\documentclass[
  aps,
  prb,
  10pt,
  reprint,
  twocolumn,
  final,
  amsfonts,
  amssymb,
  amsmath,
  hypertext,
  superscriptaddress
]{revtex4-2}
\pdfoutput=1

\usepackage{xcolor,graphicx,hyperref}
\usepackage[inline]{enumitem}
\usepackage[version=3]{mhchem}
\usepackage{bm}
\usepackage{amssymb,amsmath}
\usepackage{dcolumn}
\usepackage{physics}
\usepackage{siunitx}
\usepackage{subfigure}
\usepackage{booktabs}
\usepackage{multirow}

\ifdefined\isDraft

  \newenvironment{modify}{\color{red!80!black}}{\color{black}}
\else

\fi

\let\colorMultipoles\relax
\ifdefined\colorMultipoles
  \newcommand{\mulQ}[1]{\colorbox[HTML]{FCF7E3}{#1}}
  \newcommand{\mulM}[1]{\colorbox[HTML]{E3E9F6}{#1}}
  \newcommand{\mulT}[1]{\colorbox[HTML]{F9E3E4}{#1}}
  \newcommand{\mulG}[1]{\colorbox[HTML]{E1EBE0}{#1}}
\else
  \let\mulQ\relax
  \let\mulM\relax
  \let\mulT\relax
  \let\mulG\relax
\fi

\newcommand*{\vk}{\vb*{k}}

\graphicspath{{figure}}

\begin{document}

\preprint{APS/123-QED}

\title{Multipole analysis of spin currents in altermagnetic MnTe}

\author{Ryosuke Hirakida}
\affiliation{Department of Physics, University of Tokyo, Bunkyo, Tokyo 113-0033, Japan}

\author{Karma Tenzin}
\affiliation{Zernike Institute for Advanced Materials, University of Groningen, Nijenborgh 3, 9747 AG Groningen, The Netherlands}

\author{Chao Chen Ye}
\affiliation{Zernike Institute for Advanced Materials, University of Groningen, Nijenborgh 3, 9747 AG Groningen, The Netherlands}

\author{Berkay Kilic}
\affiliation{Zernike Institute for Advanced Materials, University of Groningen, Nijenborgh 3, 9747 AG Groningen, The Netherlands}

\author{Carmine Autieri}
\affiliation{International Research Centre MagTop, Institute of Physics, Polish Academy of Sciences, Aleja Lotnik\'ow 32/46, PL-02668 Warsaw, Poland}
\affiliation{SPIN-CNR, UOS Salerno, IT-84084 Fisciano (SA), Italy}

\author{Jagoda Sławińska}
\affiliation{Zernike Institute for Advanced Materials, University of Groningen, Nijenborgh 3, 9747 AG Groningen, The Netherlands}


\begin{abstract}
   Altermagnets, a class of unconventional antiferromagnets where antiparallel spins are connected by combined rotational and translational symmetries, have recently emerged as promising candidates for spintronic applications, as they can efficiently generate spin currents while maintaining vanishing net magnetization. Here, we investigate charge transport and spin currents in $\alpha$-MnTe, a prototypical altermagnet, using symmetry analysis within the multipole framework and fully relativistic first-principles calculations using the Kubo formalism. Our results show that different magnetic configurations with N\'eel vectors $\hat N\parallel y$ and $\hat N\parallel x$ in MnTe induce distinct order parameters.
   This distinction gives rise to spin-momentum locking with different parities and magnetic spin Hall effects (magnetic SHEs) with different anisotropies. Strikingly, our calculations show that the combination of intrinsic spin-orbit coupling and  altermagnetic spin splitting yields a large magnetic spin Hall angle of up to 16~\% rivaling or exceeding that of heavy metals such as Pt. On the other hand, the anisotropy of the magnetic SHE provides a practical means to identify the type of order parameter. This establishes, through the powerful framework of multipoles, a general approach for studying transport phenomena that extends to a broader class of altermagnets beyond MnTe.

\end{abstract}

\maketitle

\section{Introduction}
\label{sec:introduction}

Altermagnets are a highly promising class of magnets, distinct from ferromagnets and conventional antiferromagnets, and defined based on the symmetry operations that relate magnetic atoms with antiparallel spins~\cite{vsmejkal2020crystal,vsmejkal2022beyond,vsmejkal2022emerging}. In ferromagnets, no such operation exists; in antiferromagnets, these atoms are connected by inversion or translation; while in altermagnets they are connected by rotational or combined rotational–translational symmetries~\cite{vsmejkal2022beyond}. This property enables odd-rank magnetic multipoles to serve as order parameters, giving rise to parity-even spin–momentum locking in electronic states~\cite{hayami2019momentum,hayami2020bottom,yuan2021prediction,vsmejkal2022beyond}. The anisotropic spin–momentum locking has been identified as the microscopic origin of various unconventional responses to electromagnetic fields and strain~\cite{ahn2019antiferromagnetism,mazin2021prediction,vsmejkal2022giant,bhowal2022magnetic,mazin2023altermagnetism,fakhredine2023interplay,cui2023giant,cui2023efficient,fernandes2024topological,hariki2024x,hodt2024spin}, including the anomalous Hall effect (AHE)~\cite{naka2020anomalous,gonzalez2023spontaneous,kluczyk2024coexistence,takagi2025spontaneous}, piezomagnetism~\cite{aoyama2024piezomagnetic,naka2025nonrelativistic}, elasto-AHE~\cite{takahashi2025elasto}, and, most notably, the magnetic spin Hall effect (SHE)~\cite{naka2019spin,gonzalez2021efficient}.

Unlike the intrinsic time-reversal ($\mathcal{T}$) even SHE, which is limited by the strength of spin–orbit coupling (SOC), the magnetic ($\mathcal{T}$-odd) SHE is a dissipation-dependent response enabled by time-reversal symmetry breaking and exchange interactions~\cite{vzelezny2017spin,kimata2019magnetic}, and it arises from Fermi surface shifts driven by an applied electric field. Importantly, it can generate spin currents of large magnitude propagating over extended length scales. In altermagnets, the magnetic SHE is directly tied to the symmetry of magnetic multipoles, giving rise to highly anisotropic transport signatures. The key challenge, however, is to establish how specific multipolar order parameters govern the magnitude and anisotropy of the magnetic SHE, a question that remains largely unexplored and is central to advancing spin-orbitronics based on altermagnets.


In this work, we perform first-principles calculations of the prototypical altermagnet $\alpha$-MnTe, considering configurations with N\'eel vectors along the $y$- and $x$-axes. $\alpha$-MnTe has been extensively studied~\cite{ye2025dominant,belashchenko2025giant}, including the angle-resolved photoemission spectroscopy (ARPES) observations of its electronic structure~\cite{lee2024broken,krempasky2024altermagnetic,osumi2024observation} and reports of AHE~\cite{gonzalez2023spontaneous,kluczyk2024coexistence}. Here, we study the charge transport as well as $\mathcal{T}$-even and $\mathcal{T}$-odd spin Hall effects, establishing the connection between magnetic anisotropy and spin transport properties in this material. While the spin Hall response in MnTe with $\hat N \parallel y$ has been discussed recently~\cite{jeong2025magnetic}, we analyze here the distinct origins of the $\mathcal{T}$-odd SHE in $\alpha$-MnTe with $\hat N\parallel y$ and $\hat N\parallel x$. In particular, $\alpha$-MnTe with $\hat N\parallel y$ allows weak ferromagnetic ordering, whereas $\alpha$-MnTe with $\hat N\parallel x$ does not, instead realizing an altermagnetic phase that arises from a higher-rank magnetic multipole. This distinction is directly reflected in the fact that $\alpha$-MnTe with $\hat N\parallel y$ exhibits both AHE and magnetic SHE, while $\alpha$-MnTe with $\hat N\parallel x$ shows magnetic SHE but not AHE. Importantly, we identify a large magnetic spin Hall angle (SHA) in $\alpha$-MnTe, reaching 16~\% , more than twice that of Pt and comparable to or greater than other widely used spin Hall materials.

The paper is organized as follows. In Sec.~\ref{sec:computational_method}, we present the details of our first-principles calculations and the structural properties of $\alpha$-MnTe. In Sec.~\ref{sec:dft_results}, we perform a symmetry analysis based on the magnetic point group, determine the symmetry-imposed forms of the linear response tensors associated with altermagnetic order parameters, and report the calculated response coefficients. In Sec.~\ref{sec:discussion}, we propose a practical method to identify the order parameters using the AHE and magnetic SHE, and we analyze the charge-to-spin conversion efficiency in $\alpha$-MnTe. The conclusions are summarized in Sec.~\ref{sec:conclusions}.

\begin{figure*}[!htp]
    \centering
    \includegraphics[width=1.0\linewidth]{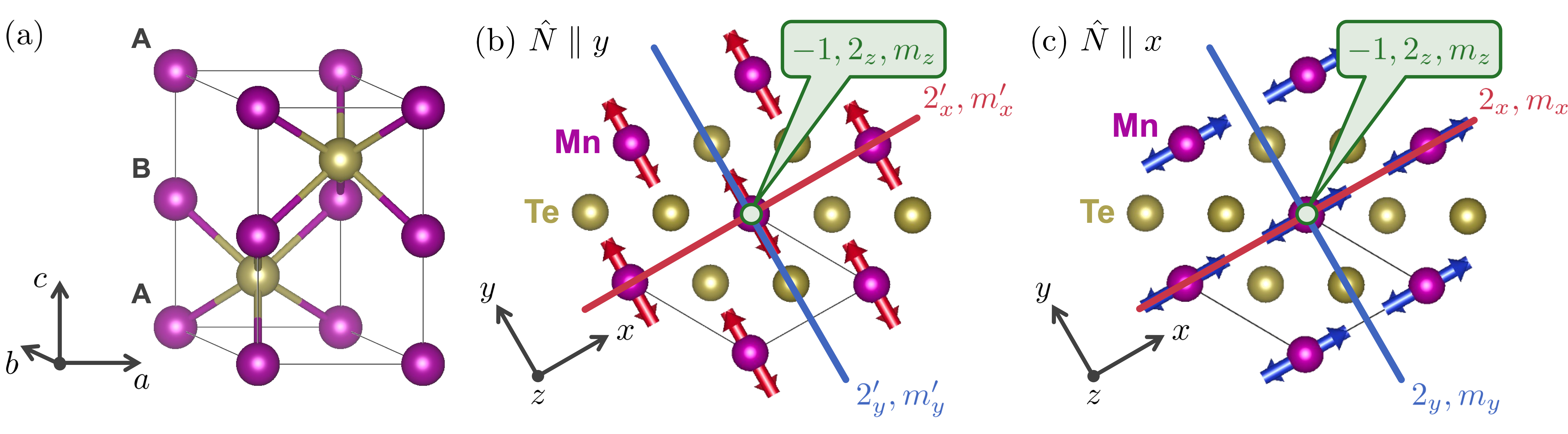}
    \caption{
        Crystal and magnetic structure of $\alpha$-MnTe~\cite{momma2011vesta}. (a) Crystal structure, (b) magnetic structure for $\hat N\parallel y$, and (c) magnetic structure for $\hat N\parallel x$. The irreducible symmetry operations of these magnetic structures are illustrated in (b), (c).
    }
    \label{fig:1_structure}
\end{figure*}

\section{Computational Method}
\label{sec:computational_method}

\subsection{Crystal and magnetic structure}

Figure~\ref{fig:1_structure} shows the crystal and magnetic structure of $\alpha$-MnTe which exhibits an altermagnetic configuration below $T_{\rm N}\sim 310$~K \cite{kriegner2017magnetic}.
In this study, we consider altermagnetic structures with N\'eel vectors oriented along the $y$- and $x$-axes. As shown in Fig.~\ref{fig:1_structure}(b), for the case of a N\'eel vector along the $y$-axis, the magnetic moments reside around the Mn atoms, pointing along the $+y$-direction in layer A and the $-y$-direction in layer B.
The corresponding magnetic point group is $m'm'm$ (\#8.4.27), and the magnetic space group is $Cm'c'm$ (\#63.462)~\cite{gonzalez2023spontaneous,kluczyk2024coexistence}.

On the other hand, for the case of a N\'eel vector along the $x$-axis, shown in Fig.~\ref{fig:1_structure}(c), the magnetic moments point along the $\pm x$-directions.
The corresponding magnetic point group is $mmm$ (\#8.1.24), and the magnetic space group is $Cmcm$ (\#63.457)~\cite{gonzalez2023spontaneous,osumi2024observation}. The irreducible symmetry operations of these magnetic point groups are summarized in Fig.~\ref{fig:1_structure}(b), (c)~\cite{aroyo2006bilbao,aroyo2006bilbaoii,ye2025dominant}.

\subsection{DFT calculations}

First-principles calculations based on density functional theory (DFT) were performed using the Vienna \textit{Ab initio} Simulation Package (VASP)~\cite{kresse1996efficient}. We used the Perdew-Burke-Ernzerhof generalized gradient approximation (PBE-GGA) for the exchange-correlation functional and projector augmented wave (PAW) pseudopotentials~\cite{blochl1994projector,perdew1996generalized}. The plane-wave energy cut-off was set to 350~eV, and the Brillouin zone was sampled using a $\Gamma$-centered $20 \times 20 \times 12$ $k$-point mesh. SOC was taken into account self-consistently in all the calculations. To accurately treat the localized Mn-$3d$ orbitals, we adopted the DFT+$U$ method~\cite{anisimov1991band,anisimov1997first} with $U = 4.0$~eV and $J = 0.97$~eV~\cite{kriegner2017magnetic}. The calculations were carried out using the experimental lattice parameters of hexagonal $\alpha$-MnTe $a = 4.134$~\AA\ and $c = 6.652$~\AA~\cite{kriegner2017magnetic}.

For the evaluation of the linear response including charge conductivity and spin currents, we used the open-source Python package \textsc{paoflow}~\cite{nardelli2018paoflow,cerasoli2021advanced}, which allows the computation of these coefficients based on the output of first-principles calculations.
\textsc{paoflow} constructs tight-binding Hamiltonians by projecting the wavefunctions obtained from VASP onto a basis of pseudo-atomic orbitals (PAOs).
The following PAO basis sets were used to accurately reproduce the DFT band structures: Mn [1s, 2s, 2p, 3s, 3p, 3d, 4s, 4p, 4d, 5s] and Te [1s, 2s, 2p, 3s, 3p, 3d, 4s, 4p, 4d, 5s, 5p, 5d, 6s, 6p].
To ensure the convergence of the calculated response coefficients, the resulting tight-binding Hamiltonians were interpolated onto dense $k$-point meshes of up to $200 \times 200 \times 124$ using Fourier interpolation with zero padding~\cite{yates2007spectral}.

\begin{table*}[!htbp]
    \centering
    \caption{Summary of response types discussed in this work.}
    \label{tab:response_summary}
    \begin{tabular}{lllll}
        \toprule
        Property & Dissipation & Name & Symbol\hspace{24.75pt} & Time-reversal \\
        \midrule
        Electric conductivity (EC) & Dissipative & Dissipative EC & $\sigma^{\mathrm{(J)}}_{ij}$ & $\mathcal{T}$-even \\
                            & Non-dissipative\hspace{24.75pt} & Anomalous Hall effect (AHE)\hspace{24.75pt} & $\sigma^{\mathrm{(E)}}_{ij}$ & $\mathcal{T}$-odd \\
        \midrule
        Rashba-Edelstein effect (REE)\hspace{24.75pt} & Dissipative & Dissipative REE & $\chi^{\mathrm{(J)}}_{sj}$ & $\mathcal{T}$-even \\
                                & Non-dissipative & Magnetic REE & $\chi^{\mathrm{(E)}}_{sj}$ & $\mathcal{T}$-odd \\
        \midrule
        Spin Hall effect (SHE) & Dissipative & Magnetic SHE & $\sigma^{\mathrm{s(J)},s}_{ij}$ & $\mathcal{T}$-odd \\
                        & Non-dissipative & Intrinsic SHE & $\sigma^{\mathrm{s(E)},s}_{ij}$ & $\mathcal{T}$-even \\
        \bottomrule
    \end{tabular}
\end{table*}

\subsection{Linear response coefficients}

In this work, we evaluate three types of linear response properties: electric conductivity, Rashba-Edelstein effect (REE) \cite{ganichev_nature, analogs2023}, and spin Hall conductivity (SHC) tensors \cite{unconventional}. Each type of response is decomposed into two contributions: the dissipative contribution $\chi^{\mathrm{(J)}}$ and the non-dissipative contribution $\chi^{\mathrm{(E)}}$. Both contributions are calculated using the Kubo formalism within the constant relaxation time approximation, given by \cite{kubo1957statistical,freimuth2014spin,li2015intraband,zhang2017strong,gonzalez2024non}:

\small
\begin{align}
    \chi^{(\mathrm{J})}(E)
    =&
    -\frac{\hbar}{\pi} \sum_{\boldsymbol{k},n,m} \frac{\Gamma^2 \, \mathrm{Re} \left[ \langle u_{n\boldsymbol{k}} | \hat{A} | u_{m\boldsymbol{k}} \rangle \langle u_{m\boldsymbol{k}} | \hat{B} | u_{n\boldsymbol{k}} \rangle \right]}{\left[(E - E_{n\boldsymbol{k}})^2 + \Gamma^2\right]\left[(E - E_{m\boldsymbol{k}})^2 + \Gamma^2\right]},
    \label{eq:chi_j}\\
    \chi^{\rm (E)}(E)
    =&
    -2\hbar
    \sum_{\vk,n\ne m}^{\substack{
      E_{n\boldsymbol{k}} \le E,\\
      E_{m\boldsymbol{k}} > E
    }}
    \frac{
      \Im\left[
        \mel{u_{n\vk}}{\hat{A}}{u_{m\vk}}
        \mel{u_{m\vk}}{\hat{B}}{u_{n\vk}}
      \right]
    }{
      (E_{n\vk}-E_{m\vk})^2+\Gamma^2
    }\notag\\
    &\times\frac{
      (E_{n\vk}-E_{m\vk})^2-\Gamma^2
    }{
      (E_{n\vk}-E_{m\vk})^2+\Gamma^2
    }.
    \label{eq:chi_e}
\end{align}
\normalsize

\noindent Here, $|u_{n\boldsymbol{k}}\rangle$ and $E_{n\boldsymbol{k}}$ denote the Bloch wavefunction and the eigenenergy for band index $n$ and wave vector $\boldsymbol{k}$, respectively. $\hbar$ is the reduced Planck constant, $\Gamma$ is the scattering rate, and $E$ is the chemical potential at which the response is evaluated.

The operators $\hat{A}$ and $\hat{B}$ correspond to the physical quantities associated with the response and the external field. For electric conductivity, we set $(\hat{A}, \hat{B}) = (\hat{j}^{\rm c}_i, \hat{j}^{\rm c}_j)$, for REE, $(\hat{A}, \hat{B}) = (\hat{s}_s, \hat{j}^{\rm c}_j)$, and for SHC, $(\hat{A}, \hat{B}) = (\hat{j}^{\mathrm{s},s}_i, \hat{j}^{\rm c}_j)$. These operators are defined as follows: the charge current operator is $\hat{j}^{\rm c}_i = e\hat{v}_i = \frac{e}{\hbar} \frac{\partial \hat{H}}{\partial k_i}$, the spin operator is $\hat{s}_s = \frac{\hbar}{2}\hat{\sigma}_s$, and the spin current operator is $\hat{j}^{\mathrm{s},s}_i = \frac{1}{2}(\hat{s}_s\hat{v}_i + \hat{v}_i\hat{s}_s)$. $\hat{\sigma}_s$ ($s=x,y,z$) are the Pauli matrices. The indices $s,i,j=x,y,z$ denote the directions of spin polarization, generated electric current, and applied electric field, respectively.

The superscripts $(\mathrm{J})$ and $(\mathrm{E})$ indicate whether the response depends on dissipation. The $(\mathrm{J})$ contribution arises from current-driven responses and is explicitly dependent on the scattering rate $\Gamma$. In the limit of small $\Gamma$, the leading order behavior is proportional to $1/\Gamma$ and can be written as~\cite{freimuth2014spin,li2015intraband}:

\begin{align}
    \chi^{(\mathrm{J})}(E) =& -\frac{1}{V}\frac{\hbar}{2\Gamma} \sum_{\boldsymbol{k},n,m} \langle u_{n\boldsymbol{k}} | \hat{A} | u_{m\boldsymbol{k}} \rangle \langle u_{m\boldsymbol{k}} | \hat{B} | u_{n\boldsymbol{k}} \rangle \notag\\
    &\times \delta(E - E_{n\boldsymbol{k}}),
    \label{eq:chi_j_small_gamma}
\end{align}

\noindent where $V$ is the volume of the Brillouin zone and $\delta(E)$ is the Dirac delta function.

In contrast, the $(\mathrm{E})$ contribution corresponds to electric field-driven responses. This contribution is independent of $\Gamma$ in the limit of small $\Gamma$~\cite{freimuth2014spin,li2015intraband}:

\begin{align}
  \chi^{(\mathrm{E})}(E) &= -2\hbar \sum_{\boldsymbol{k},n \neq m}^{\substack{E_{n\boldsymbol{k}} \le E,\\ E_{m\boldsymbol{k}} > E}} \frac{\mathrm{Im} \left[ \langle u_{n\boldsymbol{k}} | \hat{A} | u_{m\boldsymbol{k}} \rangle \langle u_{m\boldsymbol{k}} | \hat{B} | u_{n\boldsymbol{k}} \rangle \right]}{(E_{n\boldsymbol{k}} - E_{m\boldsymbol{k}})^2}.
  \label{eq:chi_e_small_gamma}
\end{align}

Table~\ref{tab:response_summary} summarizes the six types of response functions discussed in this study, including their dissipative nature, conventional names, symbols, and behavior under the time-reversal symmetry.

In this study, we address both the dissipative and non-dissipative components of electric conductivity (EC), REE, and SHE. This classification allows us to distinguish between the contributions that require time-reversal symmetry breaking and those that do not. As shown in Table~\ref{tab:response_summary}, the AHE, magnetic REE, and magnetic SHE are time-reversal-odd ($\mathcal{T}$-odd) and therefore require broken time-reversal symmetry. In contrast, dissipative EC, dissipative REE, and intrinsic SHE are time-reversal-even ($\mathcal{T}$-even) and can occur regardless of the presence of time-reversal symmetry.

\section{DFT results}
\label{sec:dft_results}

\subsection{Multipoles and spin-polarized Fermi surface}

\begin{table}[tbp]
    \caption{Active multipoles in $\alpha$-MnTe with the N\'eel vector along the $y$- and $x$-axes, classified by azimuthal quantum number $l$.}
    \label{tab:multipoles}
    \centering
    \begin{tabular}{@{}lcc@{}}
        \toprule
        & \multicolumn{2}{c}{Active multipoles in $\alpha$-MnTe} \\ \cmidrule(l){2-3}
        $l$ & $\hat N\parallel y$ & $\hat N\parallel x$ \\ \midrule
        0 & \mulQ{$Q_0$} & \mulQ{$Q_0$}, \mulT{$T_0$} \\
        1 & \mulM{$M_z$} & None \\
        2\hspace{17.8pt} & \mulQ{$Q_u, Q_v$}, \mulT{$T_{xy}$} & \hspace{17.8pt}\mulQ{$Q_u, Q_v$}, \mulT{$T_u, T_v$}\hspace{17.8pt} \\
        3 & \hspace{17.8pt}\mulM{$M_z^\alpha, M_z^\beta$}, \mulG{$G_{xyz}$}\hspace{17.8pt} & \mulM{$M_{xyz}$}, \mulG{$G_{xyz}$} \\ \bottomrule
    \end{tabular}
\end{table}

We derive the active multipoles in $\alpha$-MnTe ($\hat N\parallel y, x$) using the multipole framework~\cite{hayami2018microscopic,hayami2018classification,kusunose2020complete,hayami2020bottom,yatsushiro2021multipole,hayami2024unified, winkler2025} to reveal the spin-momentum locking in momentum space and to investigate their contribution to the response tensors. The multipoles that characterize a given system can be obtained by identifying the multipoles belonging to the totally symmetric representation of the magnetic point group of the system~\cite{yatsushiro2021multipole}. Table~\ref{tab:multipoles} lists the active multipoles in $\alpha$-MnTe for azimuthal quantum numbers $l = 0, 1, 2, 3$, derived based on its magnetic point group $m'm'm$ ($\hat N\parallel y$) and $mmm$ ($\hat N\parallel x$). Here, $Q_0$ and $T_0$ denotes the electric and magnetic toroidal monopole, $M_z$ the magnetic dipole along the $z$-axis, $Q_u$ and $Q_v$ the electric quadrupoles, $T_{xy}$, $T_u$ and $T_v$ the magnetic toroidal quadrupoles, $M_z^\alpha$, $M_z^\beta$ and $M_{xyz}$ the magnetic octupoles, and $G_{xyz}$ the electric toroidal octupole.

\begin{figure*}[!htbp]
    \centering
    \includegraphics[width=0.87\linewidth]{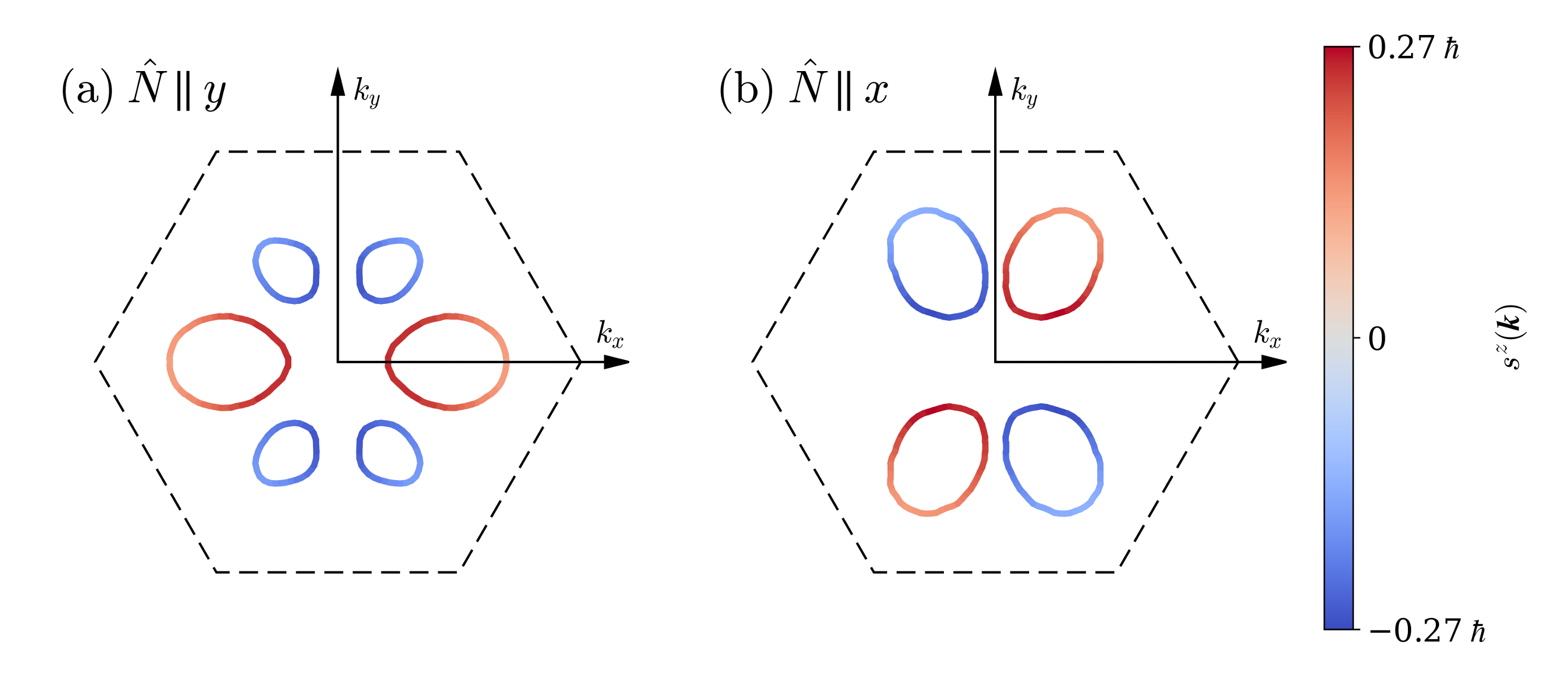}
    \caption{
        Spin-momentum locking of $\alpha$-MnTe, where the spin polarization along the $z$-axis $s^z(\boldsymbol{k})$ is projected onto the $k_xk_y$-plane at $E-E_{\rm F}=-0.1$~eV for (a) $\hat N\parallel y$ and (b) $\hat N\parallel x$. Here, the dashed hexagon represents half the size of the Brillouin zone.
    }
    \label{fig:2_fermisurf}
\end{figure*}

$\alpha$-MnTe ($\hat N\parallel y$) possesses a finite magnetic dipole $M_z$, as well as magnetic octupoles $M^\alpha_z$, $M^\beta_z$. This implies that the material can share the same symmetry as a ferromagnet magnetized along the $z$-axis, without any symmetry reduction. On the other hand, $\alpha$-MnTe ($\hat{N}\parallel x$) does not possess a magnetic dipole as its order parameter. Instead, the lowest-rank altermagnetic multipole is the magnetic octupole $M_{xyz}$. Another distinctive feature of this system is that time-reversal operation is not included among its symmetry operations, which means that $\mathcal{T}$-even multipoles cannot be distinguished from their corresponding $\mathcal{T}$-odd counterparts.

These order parameters affect the spin-momentum locking on the Fermi surface. Even though MnTe is a $g$-wave altermagnet in the nonrelativistic limit, in the relativistic case considered here, we observe a different scenario. As shown in Fig.~\ref{fig:2_fermisurf}(a), the spin-momentum locking of $\alpha$-MnTe ($\hat{N}\parallel y$) corresponds to the magnetic dipole and magnetic octupole of this system, and the $s^z$ component of the spin polarization exhibits an $s$+$d$-wave symmetry in the $k_x-k_y$ plane. The $s$-wave component (with $Q_0$ symmetry) of the spin-momentum locking arises from $M_z$, while the $d$-wave components originate from $M^\alpha_z=3\left(\tfrac{1}{2}Q_uM_z-Q_{zx}M_x-Q_{yz}M_y\right)$ and $M^\beta_z=\sqrt{15}\left(\tfrac{1}{2}Q_vM_z+Q_{zx}M_x-Q_{yz}M_y\right)$. However, since the $s$-wave component is quite small, the most dominant contribution comes from $M^\beta_z$, and thus the spin-momentum locking with $Q_v$ symmetry, the multipole coupled with $M_z$ in $M^\beta_z$, is particularly pronounced.

In contrast, as shown in Fig.~\ref{fig:2_fermisurf}(b), the spin-momentum locking of $\alpha$-MnTe ($\hat{N}\parallel x$) reflects the active magnetic octupole. In this case, $M_{xyz}=\sqrt{15}\left(Q_{xy}M_z+Q_{yz}M_x+Q_{zx}M_y\right)$ is active, and projecting $s^z$ onto the $k_x-k_y$ plane reveals a spin-momentum locking with $Q_{xy}$ symmetry, which is the multipole coupled with $M_z$ in $M_{xyz}$.

\subsection{Symmetry-imposed shape of response tensors}

\begin{table*}[tbp]
    \caption{Symmetry-imposed shape of response tensors in $\alpha$-MnTe ($\hat N\parallel y$).}
    \label{tab:symmetry_y_mnte_ahe}
    \centering
    \begin{tabular}{ll}
    \toprule
    Property & Symmetry-imposed shape of response tensors \\
    \midrule
    AHE
        & $\phantom{^{sx}}\sigma^{\mathrm{(E)}} =
        \begin{pmatrix}
            0 & \sigma^{\mathrm{(E)}}_{xy} & 0 \\
            -\sigma^{\mathrm{(E)}}_{xy} & 0 & 0 \\
            0 & 0 & 0
        \end{pmatrix}
        \leftrightarrow
        \begin{pmatrix}
            0 & M_z & 0 \\
            - M_z & 0 & 0 \\
            0 & 0 & 0
        \end{pmatrix}$
        \vspace{.5em} \\
    \begin{tabular}{l} Magnetic\\SHE \end{tabular}
        & $\sigma^{\mathrm{s(J)},x} =
        \begin{pmatrix}
            0 & 0 & \sigma^{\mathrm{s(J)},x}_{xz} \\
            0 & 0 & 0 \\
            \sigma^{\mathrm{s(J)},x}_{zx} & 0 & 0
        \end{pmatrix}
        \leftrightarrow
        \begin{pmatrix}
            0 & 0 & M''_z + 2M_z - 2T_{xy} - 2M^\alpha_z + 2M^\beta_z \\
            0 & 0 & 0 \\
            -M'_z + T'_{xy} - 3M_z + T_{xy} - 2M^\alpha_z + 2M^\beta_z & 0 & 0
        \end{pmatrix}$
        \vspace{.5em} \\
        & $\sigma^{\mathrm{s(J)},y} =
        \begin{pmatrix}
            0 & 0 & 0 \\
            0 & 0 & \sigma^{\mathrm{s(J)},y}_{yz} \\
            0 & \sigma^{\mathrm{s(J)},y}_{zy} & 0
        \end{pmatrix}
        \leftrightarrow
        \begin{pmatrix}
            0 & 0 & 0 \\
            0 & 0 & M''_z + 2M_z + 2T_{xy} - 2M^\alpha_z - 2M^\beta_z \\
            0 & -M'_z - T'_{xy} - 3M_z - T_{xy} - 2M^\alpha_z - 2M^\beta_z & 0
        \end{pmatrix}$
        \vspace{.5em} \\
        & $\sigma^{\mathrm{s(J)},z} =
        \begin{pmatrix}
            \sigma^{\mathrm{s(J)},z}_{xx} & 0 & 0 \\
            0 & \sigma^{\mathrm{s(J)},z}_{yy} & 0 \\
            0 & 0 & \sigma^{\mathrm{s(J)},z}_{zz}
        \end{pmatrix}$
        \vspace{.5em} \\
        & $\phantom{\sigma^{\mathrm{s(J)},z}} \leftrightarrow
        \begin{pmatrix}
            M'_z - T'_{xy} - 3M_z + T_{xy} - 2M^\alpha_z + 2M^\beta_z & 0 & 0 \\
            0 & M'_z + T'_{xy} - 3M_z - T_{xy} - 2M^\alpha_z - 2M^\beta_z & 0 \\
            0 & 0 & M''_z - 4M_z + 4M^\alpha_z
        \end{pmatrix}$
        \vspace{.5em} \\
    \bottomrule
    \end{tabular}
\end{table*}

\begin{table*}[tbp]
    \caption{Symmetry-imposed shape of response tensors in $\alpha$-MnTe ($\hat N\parallel x$).}
    \label{tab:symmetry_x_mnte_ahe}
    \centering
    \begin{tabular}{ll}
    \toprule
    Property\hspace{32.68pt} & Symmetry-imposed shape of response tensors \\
    \midrule
    AHE
        & $\phantom{^{sx}}\sigma^{\mathrm{(E)}}_{ij} = 0 \quad (i,j = x,y,z)$
        \vspace{.5em} \\
    \begin{tabular}{l} Magnetic\\SHE \end{tabular}
        & $\sigma^{\mathrm{s(J)},x} =
        \begin{pmatrix}
            0 & 0 & 0 \\
            0 & 0 & \sigma^{\mathrm{s(J)},x}_{yz} \\
            0 & \sigma^{\mathrm{s(J)},x}_{zy} & 0
        \end{pmatrix}
        \leftrightarrow
        \begin{pmatrix}
            0 & 0 & 0 \\
            0 & 0 & -T'_0 - 2T'_u + 2T_v + M_{xyz} \\
            0 & T'_0 - T'_u - T'_v + 3T_u - T_v + M_{xyz} & 0
        \end{pmatrix}$
        \vspace{.5em} \\
        & $\sigma^{\mathrm{s(J)},y} =
        \begin{pmatrix}
            0 & 0 & \sigma^{\mathrm{s(J)},y}_{xz} \\
            0 & 0 & 0 \\
            \sigma^{\mathrm{s(J)},y}_{zx} & 0 & 0
        \end{pmatrix}
        \leftrightarrow
        \begin{pmatrix}
            0 & 0 & T'_0 + 2T'_u + 2T_v + M_{xyz} \\
            0 & 0 & 0 \\
            -T'_0 + T'_u - T'_v - 3T_u - T_v + M_{xyz} & 0 & 0
        \end{pmatrix}$
        \vspace{.5em} \\
        & $\sigma^{\mathrm{s(J)},z} =
        \begin{pmatrix}
            0 & \sigma^{\mathrm{s(J)},z}_{xy} & 0 \\
            \sigma^{\mathrm{s(J)},z}_{yx} & 0 & 0 \\
            0 & 0 & 0
        \end{pmatrix}
        \leftrightarrow
        \begin{pmatrix}
            0 & -T'_0 + T'_u + T'_v + 3T_u - T_v + M_{xyz} & 0 \\
            T'_0 - T'_u + T'_v - 3T_u - T_v + M_{xyz} & 0 & 0 \\
            0 & 0 & 0
        \end{pmatrix}$
        \vspace{.5em} \\
    \bottomrule
    \end{tabular}
\end{table*}

The active multipoles derived in the previous section allow us to investigate types of linear responses expected in $\alpha$-MnTe based on the Kubo formula~\cite{hayami2018classification}. Tables ~\ref{tab:symmetry_y_mnte_ahe} and \ref{tab:symmetry_x_mnte_ahe} summarize the results of the multipole-based analysis applied to charge and spin transport properties which are related to the magnetic order, namely, AHE and magnetic SHE. The properties that are not directly linked to altermagnetism are provided in Appendix~\ref{app:dft_results_not_linked}.

Table~\ref{tab:symmetry_y_mnte_ahe} and \ref{tab:symmetry_x_mnte_ahe} represent the shape of the response tensors and the contributions from each multipole according to the notation of Ref.~\cite{yatsushiro2021multipole}. The left-hand side of $\leftrightarrow$ denotes the shape of the response tensor, where the nonzero elements are indicated by the corresponding tensor components. The right-hand side of $\leftrightarrow$ specifies the types of multipoles contributing to each tensor component, together with their anisotropies. For example, in $\alpha$-MnTe with $\hat{N}\parallel y$, the nonzero elements are $\sigma^{\mathrm{(E)}}_{xy}$ and $\sigma^{\mathrm{(E)}}_{yx}=-\sigma^{\mathrm{(E)}}_{xy}$, and the order parameter contributing to $\sigma^{\mathrm{(E)}}_{xy}$ is $M_z$.
Moreover, magnetic SHC in Table~\ref{tab:symmetry_y_mnte_ahe} and \ref{tab:symmetry_x_mnte_ahe} contains contributions from multipoles without superscripts, with a single prime, and with a double prime. These correspond to quadrupole-like, dipole-like, and monopole-like contributions, respectively~\cite{hayami2018classification}.

We note that AHE emerges in the $xy$-plane for $\alpha$-MnTe with $\hat N\parallel y$, whereas it vanishes for $\alpha$-MnTe with $\hat N\parallel x$. This is because the AHE shares the same symmetry as the magnetization, and the presence or absence of the magnetic dipole directly determines its existence.

The magnetic SHE of $\alpha$-MnTe with $\hat N\parallel y$ appears only in specific components, such as $\sigma^{\mathrm{s(J)},x}_{xz}$, $\sigma^{\mathrm{s(J)},x}_{zx}$, and $\sigma^{\mathrm{s(J)},z}_{xx}$. In this case, the contributions from the magnetic SHC do not overlap with those from the intrinsic SHC listed in Table VII and shown in Fig.~\ref{fig:8_ishc}(a)--(c)), which highlights their distinct physical origins. On the other hand, in $\alpha$-MnTe with $\hat N\parallel x$, the magnetic SHC appears in the same direction as the intrinsic SHC (see Fig.~\ref{fig:8_ishc}(d)--(f)). This is because in this material the $\mathcal{T}$-even multipoles cannot be distinguished from the corresponding $\mathcal{T}$-odd multipoles. The difference in the anisotropy of the magnetic SHE between the $\hat{N}\parallel y$ and $\hat{N}\parallel x$ configurations reflects the difference in the active multipoles that manifest as order parameters: $M^\alpha_z$ and $M^\beta_z$ for $\hat{N}\parallel y$, and $M_{xyz}$ for $\hat{N}\parallel x$.


\subsection{$\Gamma$ dependence}

\begin{figure*}[!htp]
    \centering
    \includegraphics[width=1.0\linewidth]{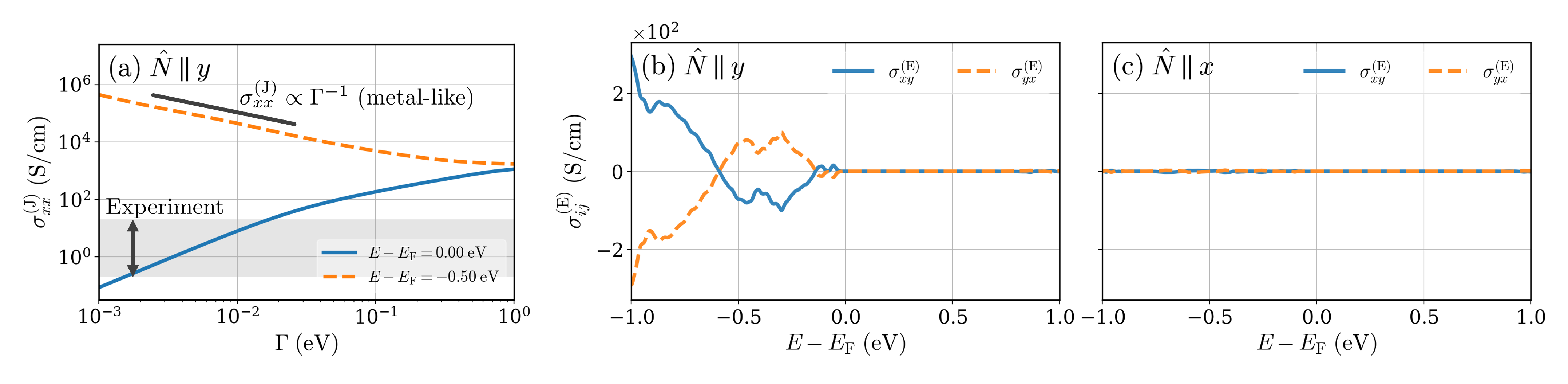}
    \caption{
        (a) Scattering rate $\Gamma$ dependence of the dissipative electric conductivity $\sigma^{\mathrm{(J)}}_{xx}$ at $E - E_{\mathrm{F}} = 0.0$~eV and $-0.5$~eV. The black solid line indicates the slope for $\sigma^{\mathrm{(J)}}_{xx}\propto\Gamma^{-1}$, while the gray shaded area represents the range of electric conductivity observed in experiments.
        Anomalous Hall conductivity $\sigma^{\mathrm{(E)}}_{ij}$ as a function of $E - E_{\mathrm{F}}$ for (b) $\hat{N}\parallel y$ and (c) $\hat{N}\parallel x$.
    }
    \label{fig:3_cond_ahc}
\end{figure*}

\begin{figure*}[htbp]
    \centering
    \includegraphics[width=1.0\linewidth]{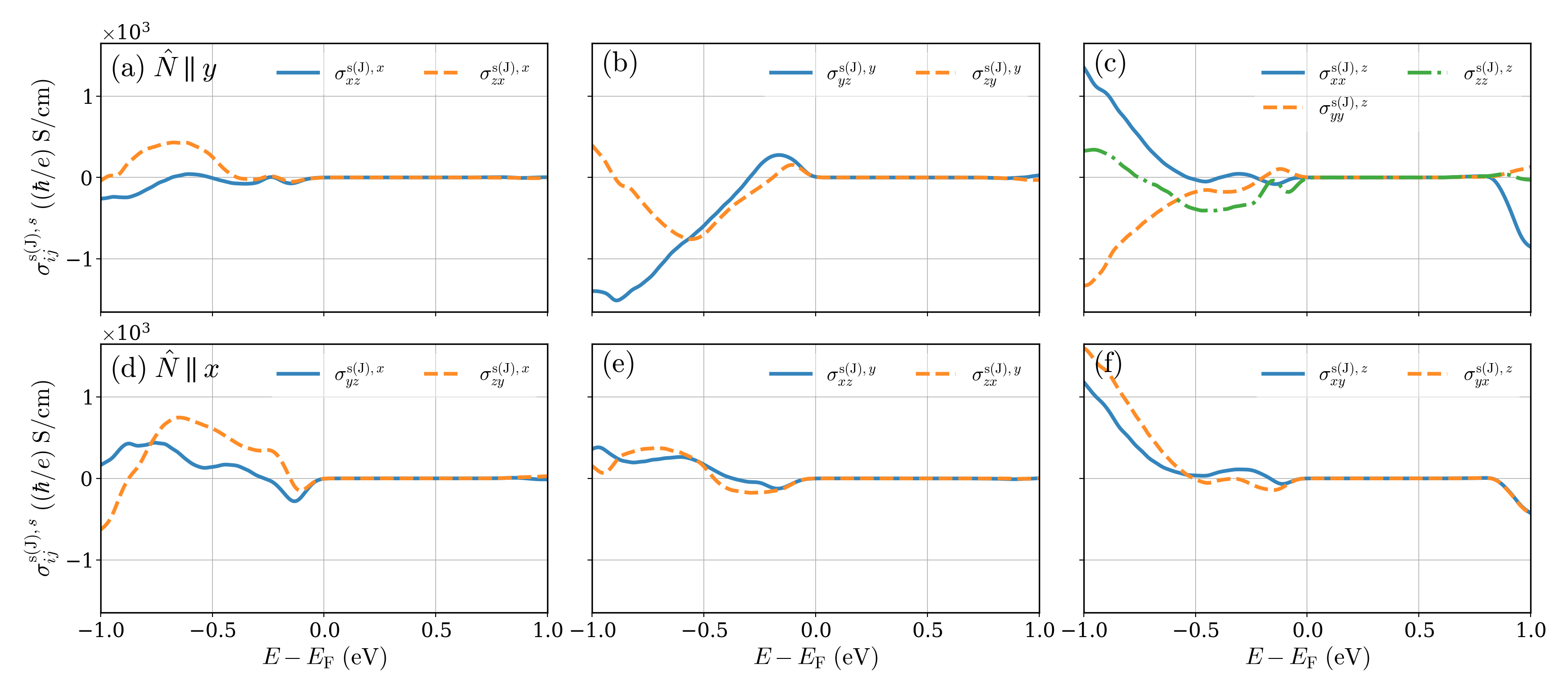}
    \caption{
        Magnetic SHC $\sigma^{\mathrm{s(J)},s}_{ij}$ as a function of $E - E_{\mathrm{F}}$. (a), (b), (c) correspond to $s=x,y,z$ with $\hat N\parallel y$, while (d), (e), (f) correspond to $s=x,y,z$ with $\hat N\parallel x$, respectively.
    }
    \label{fig:4_mshc}
\end{figure*}

In order to determine the value of the scattering rate $\Gamma$ used in the following calculations, we first evaluate the $\Gamma$-dependence of the electric conductivity. Figure~\ref{fig:3_cond_ahc}(a) shows the dissipative electric conductivity $\sigma^{\mathrm{(J)}}_{xx}$ as a function of the scattering rate $\Gamma$ for $\hat N\parallel y$. $\hat N\parallel x$ shows the same $\Gamma$ dependence. At $E - E_{\mathrm{F}} = 0$~eV, $\sigma^{\mathrm{(J)}}_{xx}$ increases as $\Gamma$ increases. This indicates that the electric conductivity of $\alpha$-MnTe behaves like an insulator. Since the Fermi energy lies at the top of the valence bands, few states contribute to conduction when $\Gamma$ is small. As $\Gamma$ increases, the number of states contributing to conduction increases due to the broadening of the spectral function, leading to an enhancement of the conductivity.

Considering this result together with the experimentally reported values of $\sigma^{\mathrm{(J)}}_{xx}$, ranging from $0.2$ to $20$~S/cm~\cite{gonzalez2021efficient,kluczyk2024coexistence}, we set the scattering rate to $\Gamma = 0.01$~eV in the rest of the calculations.
At $E - E_{\mathrm{F}} = 0$~eV and $\Gamma = 0.01$~eV, the conductivity $\sigma^{\mathrm{(J)}}_{xx}$ does not scale as $1/\Gamma$, indicating that $\Gamma$ cannot be regarded as sufficiently small. Therefore, Eq.~\eqref{eq:chi_j} is used for the calculation of conductivity. Similarly, in the rest of the calculations, Eq.~\eqref{eq:chi_j} is used for dissipative transport properties, while Eq.~\eqref{eq:chi_e} is used for non-dissipative transport properties.

\subsection{Anomalous Hall effect and magnetic spin Hall effect}

Figures~\ref{fig:3_cond_ahc}(b) and (c) show AHE calculated as a function of chemical potential $\sigma^{\mathrm{(E)}}_{ij}$. Although AHE in $\alpha$-MnTe has already been discussed in detail in Ref.~\cite{gonzalez2023spontaneous}, we present the results for completeness and validation of our computational approach. For $\hat{N}\parallel y$, the AHC appears only in the $xy$-plane, and its overall shape is consistent with Ref.~\cite{gonzalez2023spontaneous}. For $\hat{N}\parallel x$, AHE is zero, in accordance with the analysis presented in the previous sections.

Figure~\ref{fig:4_mshc} shows the energy dependence of the magnetic SHC $\sigma^{\mathrm{s(J)},s}_{ij}$. Several general trends can be observed for both $\hat{N}\parallel y$ and $\hat{N}\parallel x$. Although the band gap of $\alpha$-MnTe lies between 0.0 and 0.6~eV, SHC vanishes in an even larger region extending till approximately 0.8~eV; the finite magnitudes emerge only beyond this energy value. This behavior is likely due to the very small Fermi velocity and Berry curvature of the conduction bands around the K point (band structure is shown in Fig.~\ref{fig:6_band_structure}). Beyond 0.8~eV, the enhancement of magnetic SHC can be attributed to the contributions from other $k$-points.

On the other hand, large anomalous and spin Hall responses are observed below the band gap. For example, the AHC $\sigma^{\mathrm{(E)}}_{xy}$ of $\hat{N}\parallel y$ reaches $-227$~S/cm at $E - E_{\mathrm{F}} = -1.0$~eV. Similarly, the magnetic SHC $\sigma^{\mathrm{s(J)},y}_{yz}$ of $\hat{N}\parallel y$ shows a pronounced value of $-1.4\times 10^3$~S/cm near $E - E_{\mathrm{F}} = -0.8$~eV, while the magnetic SHC $\sigma^{\mathrm{s(J)},z}_{yx}$ of $\hat{N}\parallel x$ attains $1.5\times 10^3$~S/cm at $E - E_{\mathrm{F}} = -1.0$~eV. These trends suggest that hole doping in $\alpha$-MnTe yields much larger responses compared to electron doping.


\begin{table*}[!htp]
    \centering
    \caption{Relation between finite AHC components, order parameters, and spin-momentum locking.}
    \label{tab:ahc_multipoles}
    \begin{tabular}{@{}lcccc@{}}
        \toprule
        &  & \multicolumn{3}{c}{Spin-momentum locking} \\ \cmidrule(l){3-5}
        Finite AHC\hspace{34.25pt} & \hspace{34.25pt}Order parameters\hspace{34.25pt} & $s^x$ & $s^y$ & $s^z$ \\ \midrule
        $\sigma^{\rm(E)}_{yz}, \sigma^{\rm(E)}_{zy}$ & $M_x$ & $Q_0$ & \hspace{34.25pt}None\hspace{34.25pt} & \hspace{34.25pt}None\hspace{34.25pt} \\
        $\sigma^{\rm(E)}_{zx}, \sigma^{\rm(E)}_{xz}$ & $M_y$ & \hspace{34.25pt}None\hspace{34.25pt} & $Q_0$ & None \\
        $\sigma^{\rm(E)}_{xy}, \sigma^{\rm(E)}_{yx}$ & $M_z$ & None & None & $Q_0$ \\ \bottomrule
    \end{tabular}
\end{table*}

\section{Discussion}
\label{sec:discussion}

\subsection{Identifying order parameters using AHE and magnetic SHE}

In the case of $\alpha$-MnTe, it was clarified that the magnetic dipole $M_z$ is an order parameter for $\hat N\parallel y$, while the magnetic octupole $M_{xyz}$ is an order parameter for $\hat N\parallel x$, and these different ranks of order parameters give rise to differences in spin-momentum locking, AHE, and magnetic SHE. Conversely, from the presence of finite AHC or magnetic SHC components, one can deduce the types of order parameters and of the spin-momentum locking.

\paragraph*{AHE.}
The condition for a material to exhibit finite AHC components is that the system possesses a magnetic dipole as its order parameter. The contribution of each multipole to the AHC components is represented as~\cite{hayami2018classification}
\begin{equation}
    \sigma^{(\mathrm{E})} =
    \begin{pmatrix}
    0 & M_z & -M_y \\
    -M_z & 0 & M_x \\
    M_y & -M_x & 0
    \end{pmatrix}.
\end{equation}
Conversely, from the finite AHC components, one can infer magnetic dipoles as order parameters and the type of spin-momentum locking as summarized in Table~\ref{tab:ahc_multipoles}. The table
represents the relationship between the active order parameters and spin-momentum locking in the presence of finite AHC components and it can be interpreted as follows: If a material exhibits the $\sigma^{\mathrm{(E)}}_{yz}$ or $\sigma^{\mathrm{(E)}}_{zy}$ components, the order parameter is $M_x$. In this case, plotting the $s^x$ component in momentum space reveals a $Q_0$-type spin-momentum locking, whereas plotting the $s^y$ or $s^z$ components does not show spin-momentum locking.
Thus, Table~\ref{tab:ahc_multipoles} demonstrates that from finite AHC components, one can identify the active magnetic dipoles and the character of the spin-momentum locking.

\paragraph*{Magnetic SHE.}
If a material does not exhibit finite AHC components, it does not possess a magnetic dipole as its order parameter. However, if finite magnetic SHC components appear in the material, it has a magnetic octupole as its order parameter. Table~\ref{tab:mshc_multipoles} summarizes the correspondence between the finite magnetic SHC components, the order parameters of the material, and the type of spin-momentum locking in the hexagonal $D_{\rm 6h}$ group.

In particular, Table~\ref{tab:mshc_multipoles} indicates that if a material exhibits the $\sigma^{\mathrm{s(J)},x}_{xx}$ component, the order parameter is either $M_{3a}$ or $M_{3u}$, whereas when it shows the $\sigma^{\mathrm{s(J)},x}_{zz}$ component, the order parameter can be identified as $M_{3u}$ rather than $M_{3a}$.
The anisotropy of spin-momentum locking can be interpreted in the same way as in the case of the AHE; when plotting $s^x$, $s^y$, and $s^z$, the spin-momentum locking symmetries that appear are $2Q_u-Q_v$, $Q_{xy}$, and $Q_{zx}$, respectively. Thus, Table~\ref{tab:mshc_multipoles} demonstrates that, when the AHC vanishes, one can identify the active magnetic octupoles and the spin-momentum locking from the finite magnetic SHC components.

\begin{table*}[tbp]
    \centering
    \caption{Relation between finite magnetic SHC, order parameters, and spin-momentum locking in the hexagonal $D_{\rm 6h}$ group.}
    \label{tab:mshc_multipoles}
    \begin{tabular}{@{}lccccc@{}}
        \toprule
        \multicolumn{2}{l}{Finite magnetic SHC} &  & \multicolumn{3}{c}{Spin-momentum locking} \\ \cmidrule(r){1-2} \cmidrule(l){4-6}
        Overlapping terms & Different terms & \hspace{5.6pt}Order parameters\hspace{5.6pt} & $s^x$ & $s^y$ & $s^z$ \\ \midrule
        \multirow{2}{*}{$\sigma^{\mathrm{s(J)},x}_{xx}, \sigma^{\mathrm{s(J)},x}_{yy}, \sigma^{\mathrm{s(J)},y}_{xy}, \sigma^{\mathrm{s(J)},y}_{yx}$} &  & $M_{3a}$ & $Q_v$ & $Q_{xy}$ & \hspace{5.6pt}None\hspace{5.6pt} \\ \cmidrule(l){2-6}
        & \hspace{5.6pt}$\sigma^{\mathrm{s(J)},x}_{zz}, \sigma^{\mathrm{s(J)},z}_{xz}, \sigma^{\mathrm{s(J)},z}_{zx}$\hspace{5.6pt} & $M_{3u}$ & \hspace{5.6pt}$2Q_u-Q_v$\hspace{5.6pt} & $Q_{xy}$ & $Q_{zx}$ \\ \midrule
        \multirow{2}{*}{$\sigma^{\mathrm{s(J)},x}_{xy}, \sigma^{\mathrm{s(J)},x}_{yx}, \sigma^{\mathrm{s(J)},y}_{xx}, \sigma^{\mathrm{s(J)},y}_{yy}$} &  & $M_{3b}$ & $Q_{xy}$ & $Q_v$ & None \\ \cmidrule(l){2-6}
        & $\sigma^{\mathrm{s(J)},y}_{zz}, \sigma^{\mathrm{s(J)},z}_{yz}, \sigma^{\mathrm{s(J)},z}_{zy}$ & $M_{3v}$ & $Q_{xy}$ & \hspace{5.6pt}$2Q_u+Q_v$\hspace{5.6pt} & $Q_{yz}$ \\ \midrule
        \multirow{2}{*}{$\sigma^{\mathrm{s(J)},x}_{xz}, \sigma^{\mathrm{s(J)},x}_{zx}, \sigma^{\mathrm{s(J)},y}_{yz}, \sigma^{\mathrm{s(J)},y}_{zy}, \sigma^{\mathrm{s(J)},z}_{xx}, \sigma^{\mathrm{s(J)},z}_{yy}$} & $\sigma^{\mathrm{s(J)},z}_{zz}$ & $M^\alpha_z$ & $Q_{zx}$ & $Q_{yz}$ & $Q_u$ \\ \cmidrule(l){2-6}
        &  & $M^\beta_z$ & $Q_{zx}$ & $Q_{yz}$ & $Q_v$ \\ \midrule
        $\sigma^{\mathrm{s(J)},x}_{xz}, \sigma^{\mathrm{s(J)},x}_{zx}, \sigma^{\mathrm{s(J)},y}_{yz}, \sigma^{\mathrm{s(J)},y}_{zy}, \sigma^{\mathrm{s(J)},z}_{xx}, \sigma^{\mathrm{s(J)},z}_{yy}$\hspace{5.6pt} &  & $M_{xyz}$ & $Q_{yz}$ & $Q_{zx}$ & $Q_{xy}$ \\ \bottomrule
    \end{tabular}
\end{table*}

For $\alpha$-MnTe with $\hat N\parallel y$, finite AHC appear in the $xy$- and $yx$-components, indicating that the magnetic dipole $M_z$ manifests as an order parameter. This implies that the projection of $s^z$ onto momentum space exhibits an $s$-wave spin-momentum locking.
In contrast, for $\hat N\parallel x$, no finite AHC components are observed, and therefore the order parameter cannot be inferred from the AHC alone. Only by examining the finite magnetic SHC components of $\hat N\parallel x$ and analyzing their anisotropy, one can identify a magnetic octupole as an order parameter. This result implies that the projection of $s^z$ presents a $d$-wave spin-momentum locking in the momentum space.

To summarize, the combined analysis of AHE and magnetic SHE provides a useful method to identify whether the active order parameter is a magnetic dipole or a magnetic octupole, and to deduce the anisotropy of spin-momentum locking in altermagnets.

\subsection{Charge-to-spin conversion efficiency}

\begin{figure}[tbp]
    \centering
    \includegraphics[width=1.0\linewidth]{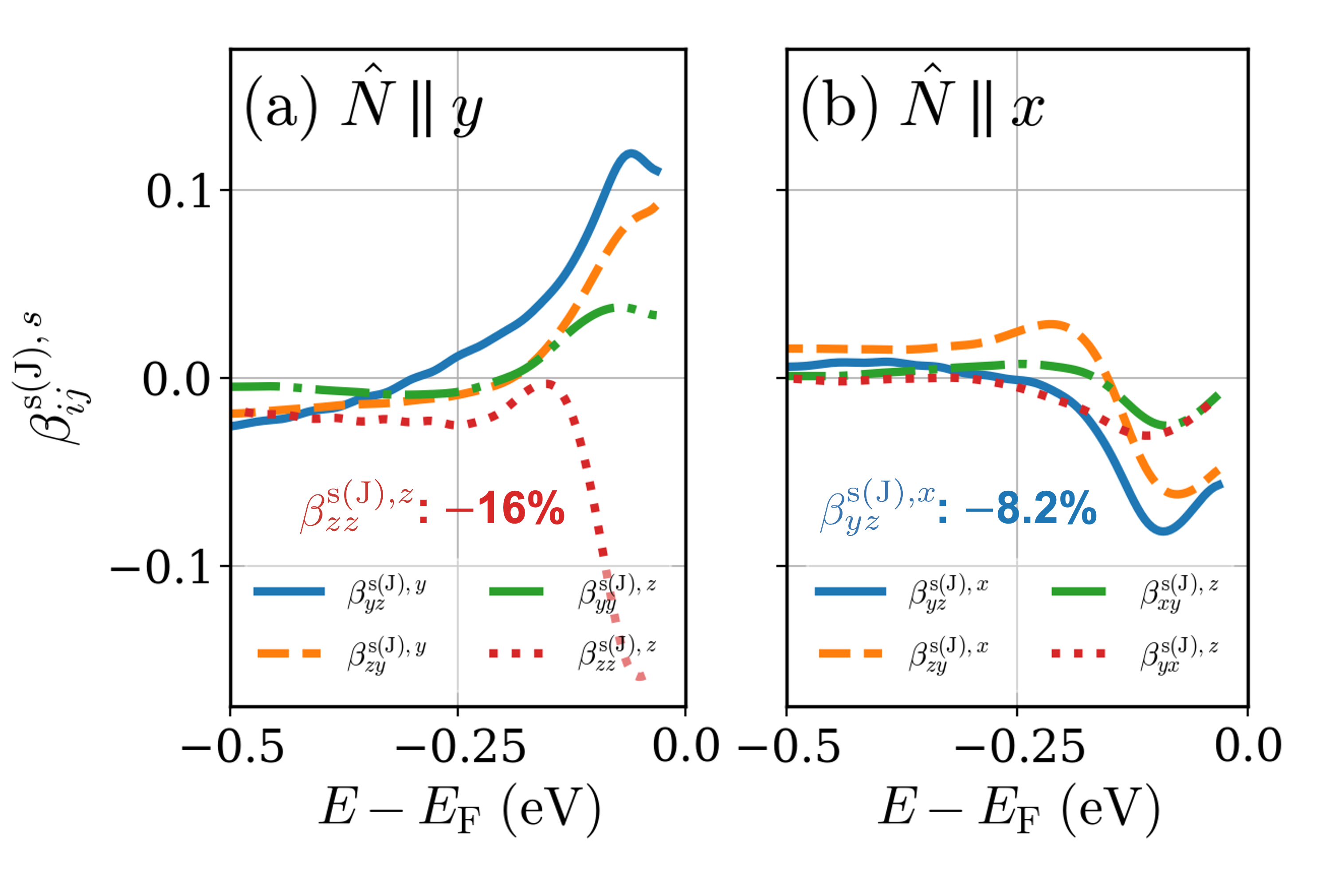}
    \caption{
        Magnetic spin Hall angle $\beta^{\mathrm{s(J)},s}_{ij}$ as a function of $E - E_{\mathrm{F}}$ for (a) $\hat N\parallel y$ and (b) $\hat N\parallel x$. $\beta^{\mathrm{s(J)},s}_{ij}$ inside the band gap is omitted because the electric conductivity is very small and gives rise to significant numerical errors.
    }
    \label{fig:5_msha}
\end{figure}

The magnitude of the magnetic SHE is commonly characterized by the dimensionless quantity known as the magnetic spin Hall angle (magnetic SHA), defined as
\begin{equation}
    \beta^{\mathrm{s(J)},s}_{ij} = \frac{e}{\hbar} \frac{\sigma^{\mathrm{s(J)},s}_{ij}}{\sigma^{\mathrm{(J)}}_{jj}}.
\end{equation}

Figure~\ref{fig:5_msha} shows $\beta^{\mathrm{s(J)},s}_{ij}$ in $\alpha$-MnTe calculated as function of chemical potential. The magnetic SHA exhibits large values near the band edges, and the maximum absolute value of $\beta^{\mathrm{s(J)},s}_{ij}$ reaches 16~\% at $E-E_{\rm F}=-0.04~{\rm eV}$ for $\hat N\parallel y$ and 8.2~\% at $E-E_{\rm F}=-0.09~{\rm eV}$ for $\hat N\parallel x$.

For $\hat N\parallel y$, the magnetic spin Hall effect is larger for the components with spin polarization along the $y$-axis ($\sigma^{\mathrm{s(J)},y}_{yz}, \sigma^{\mathrm{s(J)},y}_{zy}$) than for those with spin polarization along the $x$-axis ($\sigma^{\mathrm{s(J)},x}_{xz}, \sigma^{\mathrm{s(J)},x}_{zx}$). Similarly, for $\hat N\parallel x$, the components with spin polarization along the $x$-axis ($\sigma^{\mathrm{s(J)},x}_{yz}, \sigma^{\mathrm{s(J)},x}_{zy}$) are larger than those with spin polarization along the $y$-axis ($\sigma^{\mathrm{s(J)},y}_{xz}, \sigma^{\mathrm{s(J)},y}_{zx}$). This indicates that the components with spin polarization along the N\'eel vector are comparatively larger than the others.

The larger magnitude of SHC can be understood as follows. The magnetic SHC is determined by the Fermi velocity and the spin polarization near the Fermi surface. With respect to the Fermi velocity, no significant difference is found between $v^x_n(\boldsymbol{k})$ and $v^y_n(\boldsymbol{k})$, as suggested by the electric conductivity shown in Fig.~\ref{fig:7_cond}. However, the anisotropy of spin polarization at the Fermi surface is large. For example, in $\alpha$-MnTe with $\hat N\parallel y$, the spin polarization along the $y$-axis is significantly larger than that along the $x$-axis, as shown in the spin-projected isosurface in Ref.~\cite{ye2025dominant}. As a result, the magnetic SHC follows the relation $\sigma^{\mathrm{s(J)},y} > \sigma^{\mathrm{s(J)},x}$.

The responses associated with the ferromagnetic ordering in altermagnets have so far mainly focused on the AHE. However, the magnetic SHE also emerges due to time-reversal symmetry breaking. Our results show that the magnetic SHE in altermagnets is anisotropic, with spin Hall components polarized along the direction of the magnetic moment being particularly pronounced. These findings provide additional insight into the behavior of magnetic SHE in altermagnets and highlight another feature they share with ferromagnets.

\section{Conclusions}
\label{sec:conclusions}

In this study, we have investigated the charge and spin transport properties of $\alpha$-MnTe, a prototypical altermagnetic material, using DFT calculations combined with the symmetry analysis. We systematically derive the symmetry-imposed forms of the AHC and the magnetic SHC, and clarify how these responses are connected to the order parameters of the material. These symmetry-based predictions are further validated and quantified by numerical calculations using the Kubo formula within the constant relaxation time approximation.

Our analysis revealed that $\alpha$-MnTe with $\hat N\parallel y$ exhibits the magnetic dipole $M_z$ as its order parameter, while $\hat N\parallel x$ hosts the magnetic octupole $M_{xyz}$. These different ranks of order parameters give rise to distinct spin-momentum locking behavior: the projection of $s^z$ yields a $s$-wave-like locking in the dipole case, and a $d$-wave-like locking in the octupole case. We demonstrated that the observation of finite AHC or magnetic SHC components provides a practical route to identify the order parameters and spin-momentum locking in altermagnets.

A key finding of this work is the identification of large magnetic SHC in $\alpha$-MnTe. The magnetic SHC appears in specific tensor components, and the corresponding magnitude of magnetic SHA reaches values up to 16~\% for $\hat N\parallel y$ and 8.2~\% for $\hat N\parallel x$ near the band edges.
The magnetic SHA for $\hat N\parallel y$ is more than twice that of Pt, one of the reference spin Hall metals with SHA in the range of 5-10~\%~\cite{liu2011spin}. Although it does not surpass the exceptionally large SHAs reported for $\beta$-W (up to 30~\%)~\cite{pai2012spin}, $\alpha$-MnTe achieves comparable or even superior efficiency relative to other conventional materials such as $\beta$-Ta (12--15~\%)~\cite{liu2012spin} and AuW alloys (10~\%)~\cite{laczkowski2014experimental}.


Our results suggest that $\alpha$-MnTe is a promising material for realizing efficient charge-to-spin conversion via the magnetic SHE.
Its combination of strongly anisotropic magnetic spin Hall effect and weak net magnetization highlights the potential of altermagnets in the next-generation spintronic devices. In addition, the dissipation-driven nature of the magnetic SHE provides an alternative mechanism to the conventional, purely SOC-based effects. These findings expand opportunities for designing electronic devices that exploit the unique symmetry and transport properties of altermagnets.


\section*{Acknowledgments}
We are grateful to Masao Ogata and Masaki Kato for helpful discussions.
R.H. acknowledges the hospitality at Zernike Institute for Advanced Materials at University of Groningen, where part of this work was performed, and support by JSPS KAKENHI, Grant No. JP24KJ0580.
J.S. acknowledges the Rosalind Franklin Fellowship from the University of Groningen. J.S. and C.C.Y. acknowledge the research program “Materials for the Quantum Age” (QuMat) for financial support. This program (registration number 024.005.006) is part of the Gravitation program financed by the Dutch Ministry of Education, Culture and Science (OCW). J.S. and B.K. acknowledge the Dutch Research Council (NWO) grant OCENW.M.22.063. C.A. acknowledges the Foundation for Polish Science project “MagTop” no. FENG.02.01-IP.05-0028/23 co-financed by the European Union from the funds of Priority 2 of the European Funds for a Smart Economy Program 2021–2027 (FENG). C.A. acknowledges support from PNRR MUR project PE0000023-NQSTI. The calculations were carried out on the Dutch national e-infrastructure with the support of the SURF Cooperative using Grant No. EINF-12954.


\ifdefined\isDraft
  \bibliography{lib.bib}
\else
  \bibliography{main.bbl}
\fi


\appendix

\section{Band structures}
\label{app:band_structure}

\begin{figure*}[tbp]
    \centering
    \includegraphics[width=1.0\linewidth]{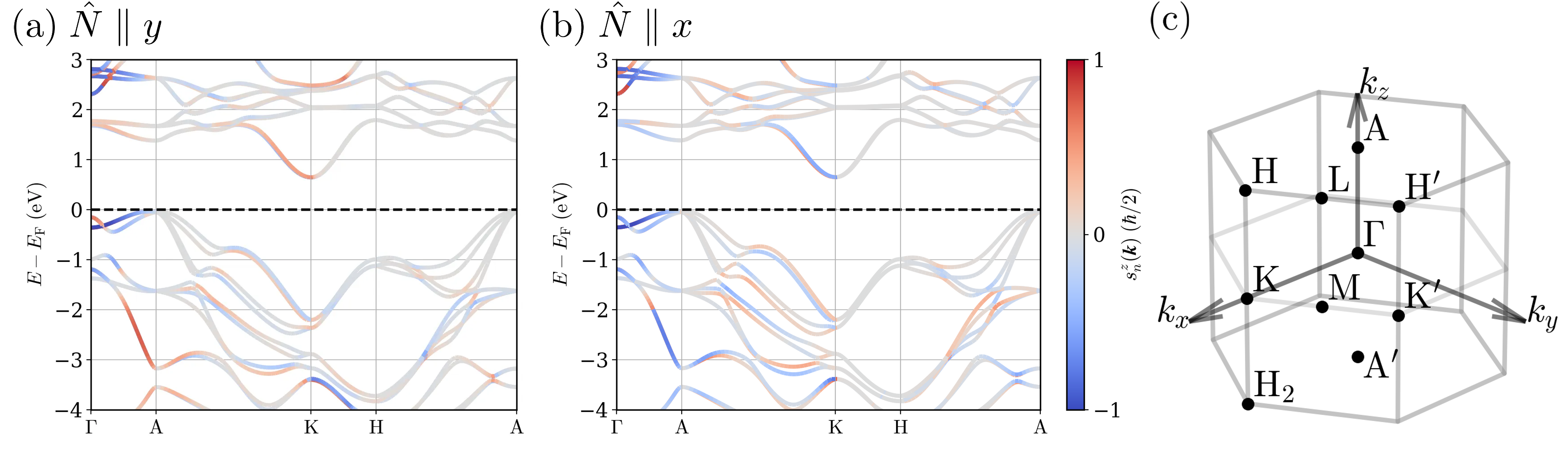}
    \caption{
        Spin-polarized band structure along the high-symmetry lines with spin polarization $s^z_n(\boldsymbol{k})$ for (a) $\hat N\parallel y$ and (b) $\hat N\parallel x$.
        (c) The corresponding first Brillouin zone and representative high-symmetry points.
        When the magnetic structure is introduced, $\alpha$-MnTe exhibits reduced symmetry and the crystal structure becomes orthorhombic. However, for the plot of the band structures, the path is taken with respect to the Brillouin zone of the parent hexagonal structure.
    }
    \label{fig:6_band_structure}
\end{figure*}

The spin-polarized band structure for $\alpha$-MnTe is shown in Fig.~\ref{fig:6_band_structure}. While the band structure for $\hat N\parallel y$ has already been reported in previous studies~\cite{krempasky2024altermagnetic,ye2025dominant}, we show it for completeness in Fig.~\ref{fig:6_band_structure}(a). The Fermi energy is set at the valence band maximum. The calculated band gap is consistent with the one reported in Ref.~\cite{faria2023sensitivity}.

\newpage
\section{Charge conductivity and intrinsic SHC}
\label{app:dft_results_not_linked}

In this appendix, we describe the transport properties that are not directly linked to the altermagnetic order, namely the dissipative EC, the REE, and the intrinsic SHE, which were not included in Sec.~\ref{sec:dft_results}.

\subsection{Symmetry-imposed shape of response tensors}

Table~\ref{tab:symmetry_y_mnte_ec} summarizes the response tensor forms, for both $\hat N \parallel y$ and $\hat N \parallel x$, as derived from the active multipoles. We note that the dissipative and magnetic REE vanish in $\alpha$-MnTe, which is a direct consequence of the inversion symmetry of the crystal. The dissipative EC and intrinsic SHE exhibit the same multipole dependence for both $\hat{N}\parallel y$ and $\hat{N}\parallel x$. This is because the symmetries of $\alpha$-MnTe with $\hat{N}\parallel y$ and $\hat{N}\parallel x$, apart from the time-reversal operation, are identical. This is different than in the magnetic SHE, where the two configurations exhibit different anisotropies as they possess different active multipoles.

The dissipative EC appears only in the longitudinal components. The three components $\sigma^{\mathrm{(J)}}_{xx}$, $\sigma^{\mathrm{(J)}}_{yy}$, and $\sigma^{\mathrm{(J)}}_{zz}$ can differ from one another, due to the contribution from the electric quadrupoles $Q_u$ and $Q_v$. The intrinsic SHC of $\alpha$-MnTe ($\hat{N}\parallel y, x$) has only conventional components, namely, it appears only when the spin polarization direction $s$, the spin current direction $i$, and the electric field direction $j$ are mutually orthogonal.

\subsection{Dissipative electric conduction and intrinsic spin Hall effect}
\begin{figure*}[tbp]
    \centering
    \includegraphics[width=1.0\linewidth]{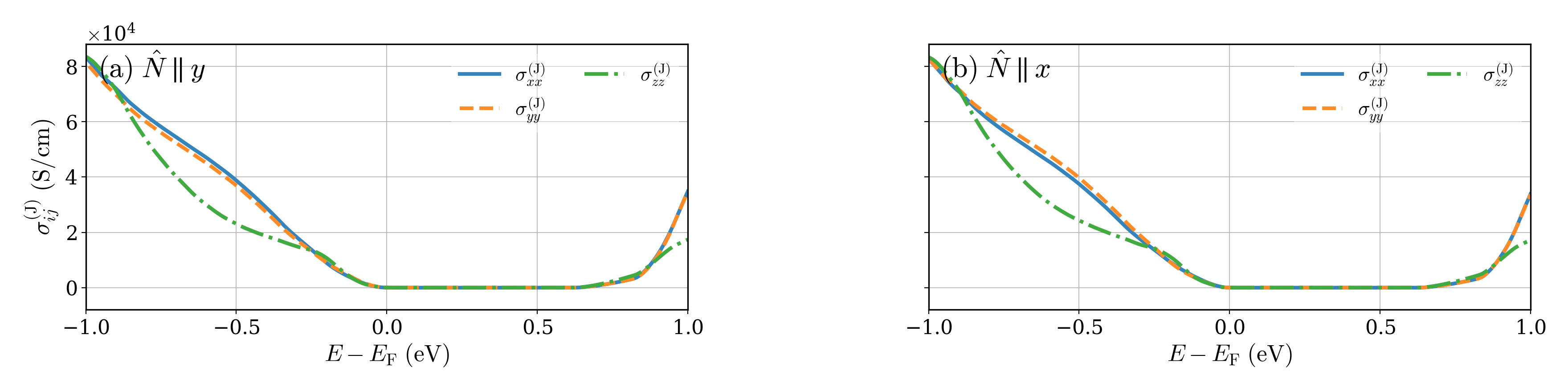}
    \caption{
        Dissipative electric conductivity $\sigma^{\mathrm{(J)}}_{ij}$ as a function of $E - E_{\mathrm{F}}$ for (a) $\hat{N}\parallel y$ and (b) $\hat{N}\parallel x$.
    }
    \label{fig:7_cond}
\end{figure*}

\begin{figure*}[tbp]
    \centering
    \includegraphics[width=1.0\linewidth]{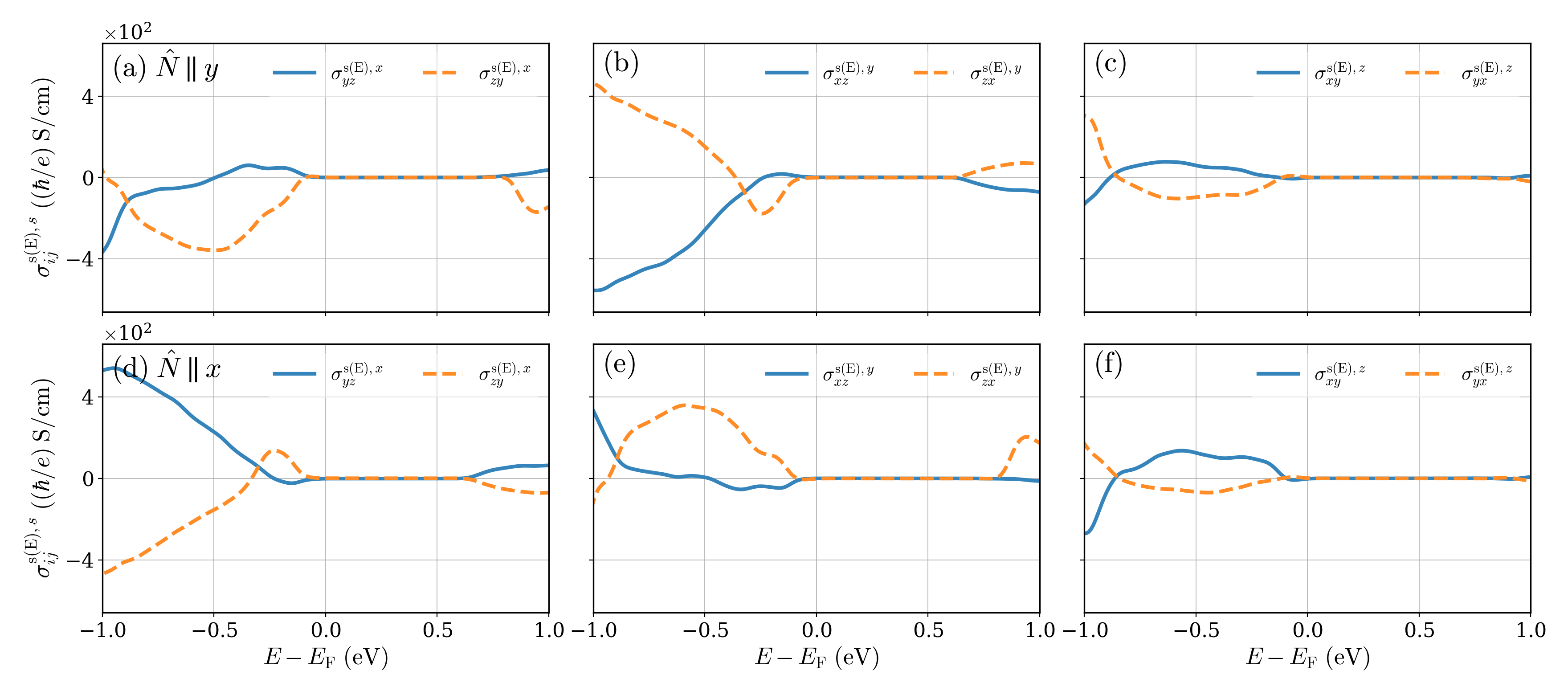}
    \caption{
        Intrinsic SHC $\sigma^{\mathrm{s(E)},s}_{ij}$ as a function of $E - E_{\mathrm{F}}$. (a), (b), (c) correspond to $s=x,y,z$ with $\hat N\parallel y$, while (d), (e), (f) correspond to $s=x,y,z$ with $\hat N\parallel x$, respectively.
    }
    \label{fig:8_ishc}
\end{figure*}

For the dissipative EC and intrinsic SHE shown in Fig.~\ref{fig:7_cond} and Fig.~\ref{fig:8_ishc}, the trends are similar to those of the AHE and magnetic SHE; hole doping produces a much stronger response than electron doping. Note that the intrinsic SHE of MnTe has six independent components of the SHC tensor, in contrast to the nonmagnetic case $P6_{3}/mmc$ (\#194), where among the six existing SHC components, only three are independent \cite{unconventional}. This is a natural consequence of the reduced symmetry of the magnetic system.

\begin{table*}[!bp]
    \caption{Symmetry-imposed shape of response tensors in $\alpha$-MnTe the same for both $\hat N\parallel y$ and $\hat N\parallel x$.}
    \label{tab:symmetry_y_mnte_ec}
    \centering
    \begin{tabular}{ll}
    \toprule
    Property\hspace{5.37pt} & Symmetry-imposed shape of response tensors \\
    \midrule
    EC
        & $\phantom{^{s,x}}\sigma^{\mathrm{(J)}} =
        \begin{pmatrix}
            \sigma^{\mathrm{(J)}}_{xx} & 0 & 0 \\
            0 & \sigma^{\mathrm{(J)}}_{yy} & 0 \\
            0 & 0 & \sigma^{\mathrm{(J)}}_{zz}
        \end{pmatrix}
        \leftrightarrow
        \begin{pmatrix}
            Q_0 - Q_u + Q_v & 0 & 0 \\
            0 & Q_0 - Q_u - Q_v & 0 \\
            0 & 0 & Q_0 + 2Q_u
        \end{pmatrix}$
        \vspace{.5em} \\
    REE
        & $\phantom{^{s,x}}\chi^{\mathrm{(J)}}_{ij} = 0 \quad (i,j = x,y,z)$
        \vspace{.5em} \\
        & $\phantom{^{sx}}\chi^{\mathrm{(E)}}_{ij} = 0 \quad (i,j = x,y,z)$
        \vspace{.5em} \\
    \begin{tabular}{l} Intrinsic\\SHE \end{tabular}
        & $\sigma^{\mathrm{s(E)},x} =
        \begin{pmatrix}
            0 & 0 & 0 \\
            0 & 0 & \sigma^{\mathrm{s(E)},x}_{yz} \\
            0 & \sigma^{\mathrm{s(E)},x}_{zy} & 0
        \end{pmatrix}
        \leftrightarrow
        \begin{pmatrix}
            0 & 0 & 0 \\
            0 & 0 & -Q'_0 - 2Q'_u + 2Q_v + G_{xyz} \\
            0 & Q'_0 - Q'_u - Q'_v + 3Q_u - Q_v + G_{xyz} & 0
        \end{pmatrix}$
        \vspace{.5em} \\
        & $\sigma^{\mathrm{s(E)},y} =
        \begin{pmatrix}
            0 & 0 & \sigma^{\mathrm{s(E)},y}_{xz} \\
            0 & 0 & 0 \\
            \sigma^{\mathrm{s(E)},y}_{zx} & 0 & 0
        \end{pmatrix}
        \leftrightarrow
        \begin{pmatrix}
            0 & 0 & Q'_0 + 2Q'_u + 2Q_v + G_{xyz} \\
            0 & 0 & 0 \\
            -Q'_0 + Q'_u - Q'_v - 3Q_u - Q_v + G_{xyz} & 0 & 0
        \end{pmatrix}$
        \vspace{.5em} \\
        & $\sigma^{\mathrm{s(E)},z} =
        \begin{pmatrix}
            0 & \sigma^{\mathrm{s(E)},z}_{xy} & 0 \\
            \sigma^{\mathrm{s(E)},z}_{yx} & 0 & 0 \\
            0 & 0 & 0
        \end{pmatrix}
        \leftrightarrow
        \begin{pmatrix}
            0 & -Q'_0 + Q'_u + Q'_v + 3Q_u - Q_v + G_{xyz} & 0 \\
            Q'_0 - Q'_u + Q'_v - 3Q_u - Q_v + G_{xyz} & 0 & 0 \\
            0 & 0 & 0
        \end{pmatrix}$
        \vspace{.5em} \\
    \bottomrule
    \end{tabular}
\end{table*}

\end{document}